\documentclass[a4paper,12pt]{paper} 
\usepackage{amsmath}
\usepackage{braket}
\usepackage{graphicx}

\title{The effect of electron-phonon interaction on the thermoelectric properties of defect zigzag nanoribbons}

\author{%
  D.V. Kolesnikov$^{1a}$,
  D.A. Lobanov$^1$,
  V.A. Osipov$^1$}

\begin{document}
\maketitle
\begin{center}
\begin{small}
  $^1$Bogoliubov Laboratory of Theoretical Physics,\\ Joint Institute for Nuclear Research,\\ 141980 Dubna, Moscow region, Russia\\
  $^a$e-mail:kolesnik@theor.jinr.ru
\end{small}
\end{center}
\begin{abstract}
Thermoelectric properties of graphene nanoribbons with periodic edge vacancies and antidot lattice are investigated. The electron-phonon interaction is taken into account in the framework of the Hubbard-Holstein model with the use of the Lang-Firsov unitary transformation scheme. The electron transmission function, the thermopower and the thermoelectric figure of merit are calculated. We have found that the electron-phonon interaction causes a decrease in the peak values of the thermoelectric figure of merit and the shift of the peak positions closer to the center of the bandgap. The effects are more pronounced for the secondary peaks that appear in the structures with periodic antidot.
\end{abstract}

\section{Introduction}
The influence of electron-phonon interaction on the properties of graphene nanostructures was widely investigated \cite{bory,manes}.  One of the important properties of the electron-phonon coupling in graphene is its non-adiabatic character \cite{non,non2}. Due to this fact, the density-density electron-phonon coupling with the graphene ZO optical mode, described by the Hubbard-Holstein Hamiltonian, is widely used in graphene \cite{stauber,mogulkoc}. In the case of graphene nanoribbons this model leads to the appearance of small polarons \cite{mog2}. Owing to the relative weakness of electron-phonon interaction in graphene \cite{cng}, it has only limited influence on the electron current. On the other hand, the thermocurrent is highly sensitive to any small variations of the electron transmission function. Therefore, the electron-phonon interaction can play an important role in thermoelectric devices based on graphene nanoribbons. 

The thermoelectric figure of merit is defined as $ZT = G_e S^2 T/\kappa,$
where $G_e$ is the electron conductance, $S$ is the Seebeck coefficient, $T$ is the temperature, and $\kappa=\kappa_e+\kappa_{ph}$ is the thermal conductance consisting of the sum of the electron $\kappa_e$ and phonon $\kappa_{ph}$ contributions. The thermoelectric figure of merit, which is determined primarily by the electronic conductivity and the Seebeck coefficient, is very sensitive to the behavior of the electron transmission function. 

The case of zigzag graphene nanoribbons (ZGNR) with periodic edge vacancies was investigated in \cite{ko}. It was shown that the periodic edge vacancies with the period equal to three elementary translations lead to a finite band gap with almost degenerate states on the valence and conduction band edges. This results in the sharp step-like behavior of the electron transmission function near band edges. Such behavior leads to the increased values of figure of merit, when the energy difference between nearest levels on the band edge is less than or equal to $3k_B T$, where $k_B$ is the Boltzmann constant. The energy of electron-phonon coupling is comparable with this difference and thermal activation energy. This makes electron-phonon interaction an important factor, which can have significant impact on thermoelectric properties of ZGNRs with edge defects.

\section{General formalism}
We consider a periodic structure, that can be arbitrarily divided into three parts: semi-infinite periodic leads (left and right) and a finite central part. In the absence of the electron-phonon interaction, the electronic Hamiltonian is written as
\begin{eqnarray}
H_0 =  \sum_{i\sigma}\epsilon_i C^\dagger_{i\sigma}C_{i\sigma} + H_1 +H_2 + H_U,\\
H_1=t\sum_{<i,j>,\sigma}[C^{\dagger}_{i\sigma}C_{j\sigma} + C^{\dagger}_{j\sigma}C_{i\sigma}],\\
H_2= t_2\sum_{<<i,j>>,\sigma}[C^{\dagger}_{i\sigma}C_{j\sigma} + C^{\dagger}_{j\sigma}C_{i\sigma}],\\
H_U=  U\sum_i C^\dagger_{i\sigma}C_{i\sigma}C^\dagger_{i,-\sigma}C_{i,-\sigma},
\end{eqnarray}
where  $ C_{i\sigma}(C^\dagger_{i\sigma})$ are the electron annihilation (creation) operators, $<i,j>$ and $<<i,j>>$ correspond to the sum over the nearest-neighbor (NN) and the next-to-nearest neighbor (NNN) atoms, $t$ and $t_2$ correspond to the NN and NNN hopping integrals, and $U$ is the electron-electron interaction strength. The Hubbard term is calculated in the mean-field approximation $U\sum_{i\sigma} n_{i\sigma}n_{i,-\sigma} \rightarrow U\sum_{i\sigma} n_{i\sigma}<n_{i-\sigma}>$, where $n_{i\sigma}=C^\dagger_{i\sigma}C_{i\sigma}$ is the electron spin density. 

To include the electron-phonon interaction, we modify the Hamiltonian by adding the phonon part and the Holstein-type electron-phonon interaction \cite{mogulkoc}
\begin{eqnarray}
H = H_0 + \sum_i \hbar \omega_0 b^\dagger_i b_i + g\sum_{i\sigma}(b^\dagger_i+b_i)n_{i\sigma},\label{Hph}
\end{eqnarray}
where $\hbar$ is the modified Planck constant, $\omega_0$ is the frequency of the dispersionless mode corresponding to the ZO mode in graphene, $b^\dagger_i$ and $b_i$ are the creation/annihilation operators, and $g$ is the electron-phonon coupling parameter. We consider the Lang-Firsov unitary transformation to diagonalize the phonon part of the Hamiltonian
\begin{equation}
U_1 = \exp\left[\frac{g}{\hbar\omega_0}\sum_{i\sigma}(b^\dagger_i-b_i)n_{i\sigma} \right],\label{U_lf}
\end{equation}
so that the phonon and electron annihilation (creation) operators are transformed as $\tilde{b}_i=b_i-\sum_\sigma n_{i\sigma} g/(\hbar \omega_0)$, $\tilde{b}^\dagger_i=b^\dagger_i-\sum_\sigma n_{i\sigma} g/(\hbar \omega_0)$, and $\tilde{C}_i=C_i X_i$, $\tilde{C}^\dagger_i=C^\dagger_i X^\dagger_i$, where $X_i=\exp[-g/(\hbar\omega_0)(b^\dagger_i-b_i)]$. We can make an ansatz for the ground state of the system $\ket{\psi}=\ket{\psi_{el}}\otimes U_1\ket{0_{ph}}$, where $\ket{0_{ph}}$ corresponds to the phonon vacuum. The transformed Hamiltonian $\tilde{H}=U_1^{-1}HU_1$ can be written as \cite{chakraborty}
\begin{eqnarray}
\tilde{H} = \sum_{i\sigma}\left( \epsilon_i - \frac{g^2}{\hbar\omega_0} \right) C^\dagger_{i\sigma}C_{i\sigma} + \tilde{H}_1 +\tilde{H}_2 + \tilde{H}_U,\\
\tilde{H}_1 = \tilde{t}\sum_{<i,j>,\sigma}[C^{\dagger}_{i\sigma}C_{j\sigma} + C^{\dagger}_{j\sigma}C_{i\sigma}],\\
\tilde{H}_2 = \tilde{t}_2\sum_{<<i,j>>,\sigma}[C^{\dagger}_{i\sigma}C_{j\sigma} + C^{\dagger}_{j\sigma}C_{i\sigma}] ,\\
\tilde{H}_U = \sum_{i\sigma}\left(U-\frac{g^2}{\hbar\omega_0} \right)C^\dagger_{i\sigma}C_{i\sigma}C^\dagger_{i,-\sigma}C_{i,-\sigma},
\end{eqnarray}
where $\tilde{t}=t\exp(-g^2/(\hbar\omega_0^2))$, $\tilde{t}_2=t_2\exp(-g^2/(\hbar\omega_0^2))$. Another effect of the electron-phonon interaction is its influence on the self-energies of leads: an ansatz of the form $\Sigma^{<}(E)=\Lambda(E)\Sigma_0^{<}(E)$ \cite{ferretti} would give us the following equations for the Green's function in the form similar to \cite{datta} 
\begin{eqnarray}
\tilde{T}_{e\sigma}(E) = \mathrm{Tr}[\Gamma_L G^r_\sigma(E) \Gamma_R\Lambda G^a_\sigma(E)], \label{negf0} \\
G^r_\sigma(E) = [I(E+i\eta) + \tilde{H}_C-\Sigma^r_L-\Sigma^r_R]^{-1},\\
\Gamma_{\alpha} = -2\,\mathrm{Im}\,\Sigma^r_{\alpha}(E),\; \alpha=L,R,\\
\Sigma^r_{\alpha} = \tilde{H}_{C\alpha}g^r_{\alpha}\tilde{H}_{\alpha C},
\label{negf}
\end{eqnarray}
where $G^r_\sigma(E)$ $(G^a_\sigma(E))$ is the retarded (advanced) Green's function of the central part, $\Gamma_{L,R}$ and $\Sigma^r_{L,R}$ are the electron broadening  and the self-energy matrix of the left(right) lead, $\tilde{H}_{C}$, $\tilde{H}_{CL}$ and $\tilde{H}_{RC}$ are the transformed Hamiltonian of the central part and blocks corresponding to the interaction between the left (right) lead and the central part, respectively, $I$ is the unit matrix, $\eta$ is a small parameter, and $g^r_{L,R}$ is the retarded surface Green's function of the left (right) lead. This function is calculated using the Sancho-Rubio scheme~\cite{LopezSancho}.
The Landauer-like formula for the spin-dependent current through the system is expressed as
\begin{equation}
I_{e\sigma} = \frac{q}{h}\int_{-\infty}^{\infty} \tilde{T}_{e\sigma}(E)( f_L-f_R )dE,\label{Iesigma}
\end{equation}
where $q$ is the electron charge, $f_{L,R}=f(E,\mu_{L,R} )$ is the Fermi-Dirac distribution function of left and right leads and $h$ is the Planck constant, so that the electron-phonon interaction results in the effective modification of the electron transmission. In the linear response regime the electron transport approaches equilibrium. In this case, the matrix $\Lambda$ can be written as $\Lambda=I\exp[-g^2/(\hbar\omega_0)^2].$

The electron current and electron thermal current through the device in the linear response regime can be defined as \cite{Jiang}
\begin{eqnarray}
I_e = \frac{2q}{h}\int_{-\infty}^\infty \tilde{T}_e(E)(f_L-f_R)dE ,\label{Ie}\\
I_Q = \frac{2}{h}\int_{-\infty}^\infty \tilde{T}_e(E)(E-\mu)(f_L-f_R)dE ,\label{Iq}
\end{eqnarray}
where $\tilde{T}_e(E)=\sum_\sigma \tilde{T}_{e\sigma}(E)/2$ is the energy-dependent effective electron transmission function.

The phonon thermal conductance can be expressed as 
\begin{equation}
\kappa_{ph} = \frac{1}{2\pi}\int_0^{\infty} T_{ph}(\omega)\hbar\omega\frac{\partial n(\omega,T)}{\partial \omega}d \omega,\label{kph}
\end{equation}
where $T_{ph}(\omega)$ is the phonon transmission function, $\omega$ is the phonon frequency, and $n(\omega,T)$ is the Bose-Einstein distribution function. The phonon transmission is calculated using the NEGF approach similar to (\ref{negf0})-(\ref{negf}), where the substitution $H\rightarrow K,\;E+i\eta\rightarrow (\omega+i\eta)^2$ is made, so that the Hamiltonian matrix $H$ is replaced with the force constant matrix $K$ (see \cite{Jiang} for detail).
We calculate the force constant matrix of the structure within the 4th nearest neighbor approximation by using a set of constants from \cite{Zimmerman}, and the influence of electron-phonon interaction on the thermal conductance is neglected. In the linear response regime, according to (\ref{negf0},\ref{Ie}-\ref{Iq}) one can find the electrical conductance, the Seebeck coefficient and the electron thermal conductance from the effective electron transmission as
\begin{eqnarray}
G_e = q^2 L_0,\;
S = \frac{L_1}{qT L_0},\;
\kappa_e = \frac{1}{T}(L_2-\frac{L_1^2}{L_0}),\label{gpisdefs}
\end{eqnarray}
where
\begin{equation}
L_n(\mu,T) =  -\frac{2}{h} \int_{-\infty}^{\infty} \tilde{T}_e(E)(E-\mu)^n \frac{df(E,\mu)}{dE}dE.\label{Ln}
\end{equation}
In the graphene-based materials, the exact behavior of the transmission function is important for the thermoelechric characteristics of the device under consideration. The important parameter would be the energy difference between the nearest energy levels on the conduction and valence band edges, which also corresponds to the "sharpness" of the electron transmission function near the band edges \cite{ko}. The electron-phonon interaction, which is relatively small in the graphene-based structures, could increase such sharpness, which would result in the significant change in the figure of merit.

\section{Results and discussion}

We have calculated the effective electron transmission, phonon and electron thermal conductance, thermopower and the thermoelectric figure of merit in the ZGNR with periodic edge vacancies and antidots, 6 and 8 atoms wide (see Fig.1). 

\begin{figure*}[htb]
\includegraphics*[width=0.7\textwidth]{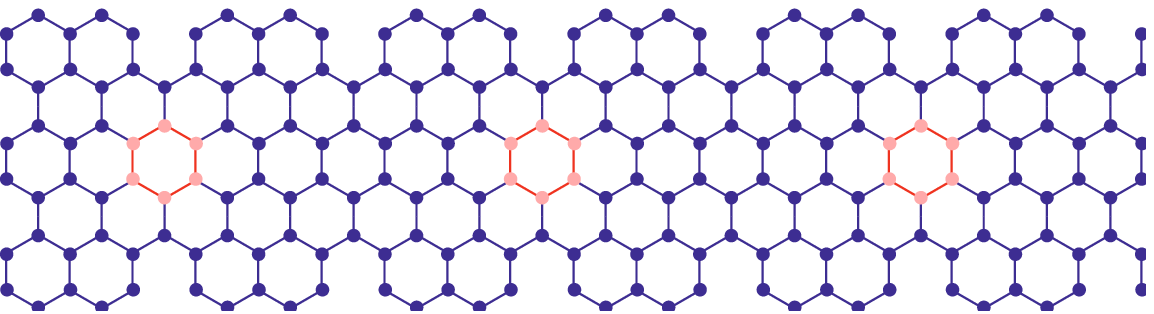}\\ \vspace{0.2cm}
\includegraphics*[width=0.7\textwidth]{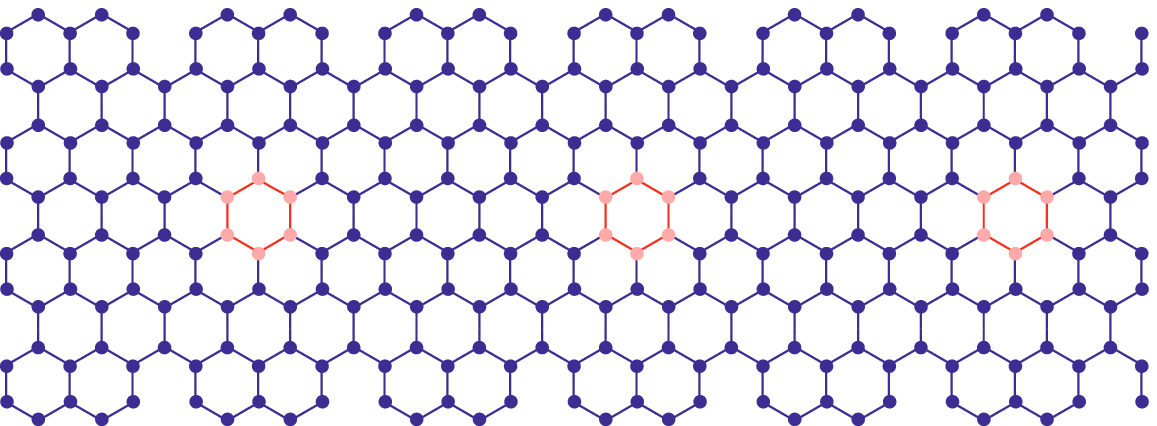} \label{fig1}
\caption{ZGNR structures, 6 (top) and 8 (bottom) atoms wide. The periodic antidot lattice is shown in red.}
\end{figure*}

We have used the nearest (NN) and next-nearest-neighbor (NNN) hopping parameters $t=2.7 \;eV$ \cite{prl} and $\;t_2=-0.1t$ \cite{Kretinin}, the electron-electron interaction strength $U=1.1 t$ and the energy of phonon mode $\hbar\omega_0=0.15$ eV, while we vary the electron-phonon coupling $g$ from 0 to 0.075 eV \cite{mogulkoc}. The thermal conductance of the structures is shown in Table \ref{t1}.
\begin{table*}
\caption{The phonon thermal conductance $\kappa_{ph}$ at T=300 K}
\label{t1}
\begin{tabular}{lllll}
\hline\noalign{\smallskip}
Structure & ZZ-6+vac.&ZZ-6+vac.+antidot& ZZ-8+vac.&ZZ-8+vac.+antidot\\
\noalign{\smallskip}\hline\noalign{\smallskip}
$\kappa_{ph}$, nW/K & 0.9917  & 0.5282  & 1.376 & 0.8556 \\
\noalign{\smallskip}\hline
\end{tabular}
\end{table*}
Figs.2 and 3 show the electronic bandstructure for the electronic Hamiltonian and the effective total electron transmission function. 
\begin{figure*}[htb]
\includegraphics*[width=0.48\textwidth]{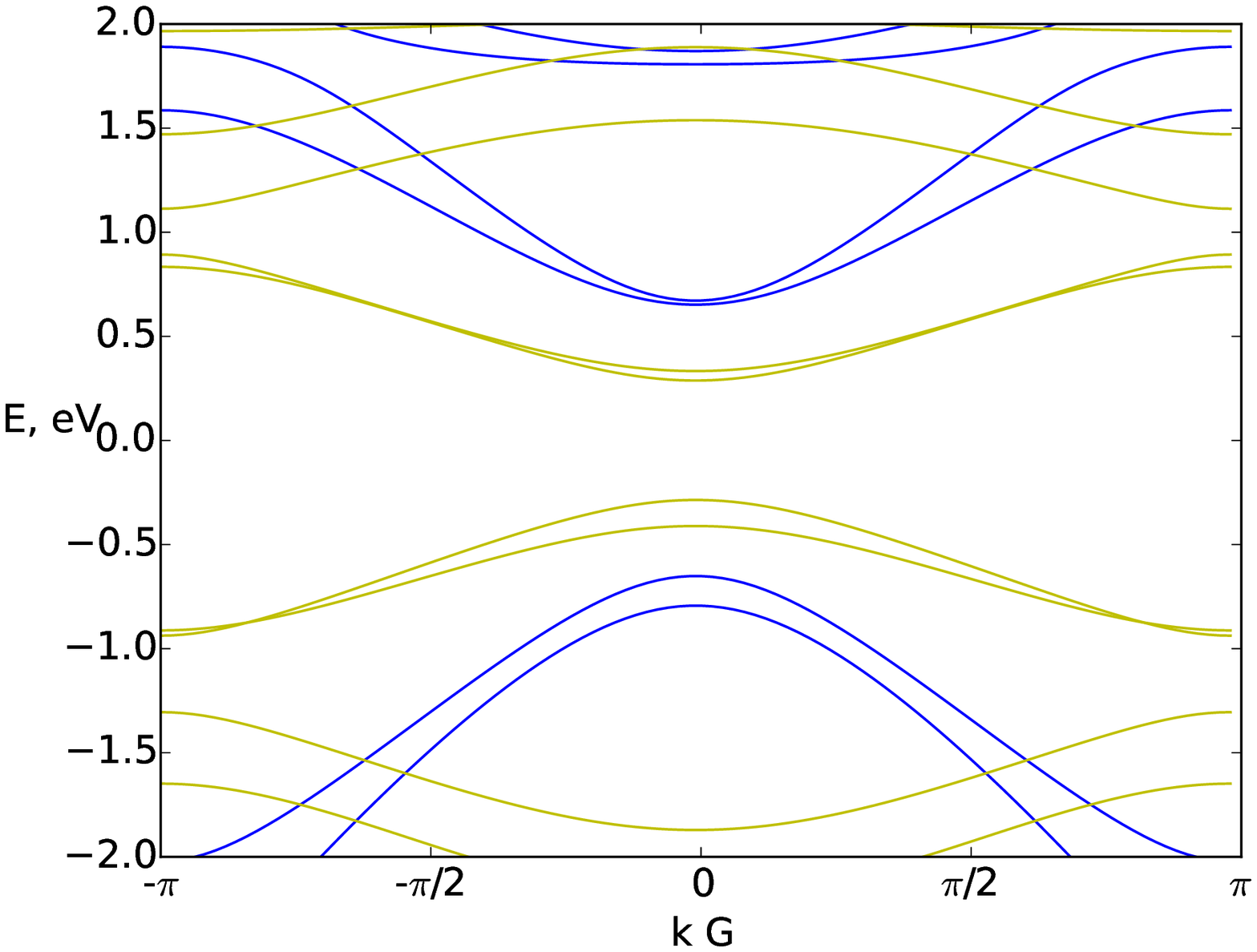}
\includegraphics*[width=0.48\textwidth]{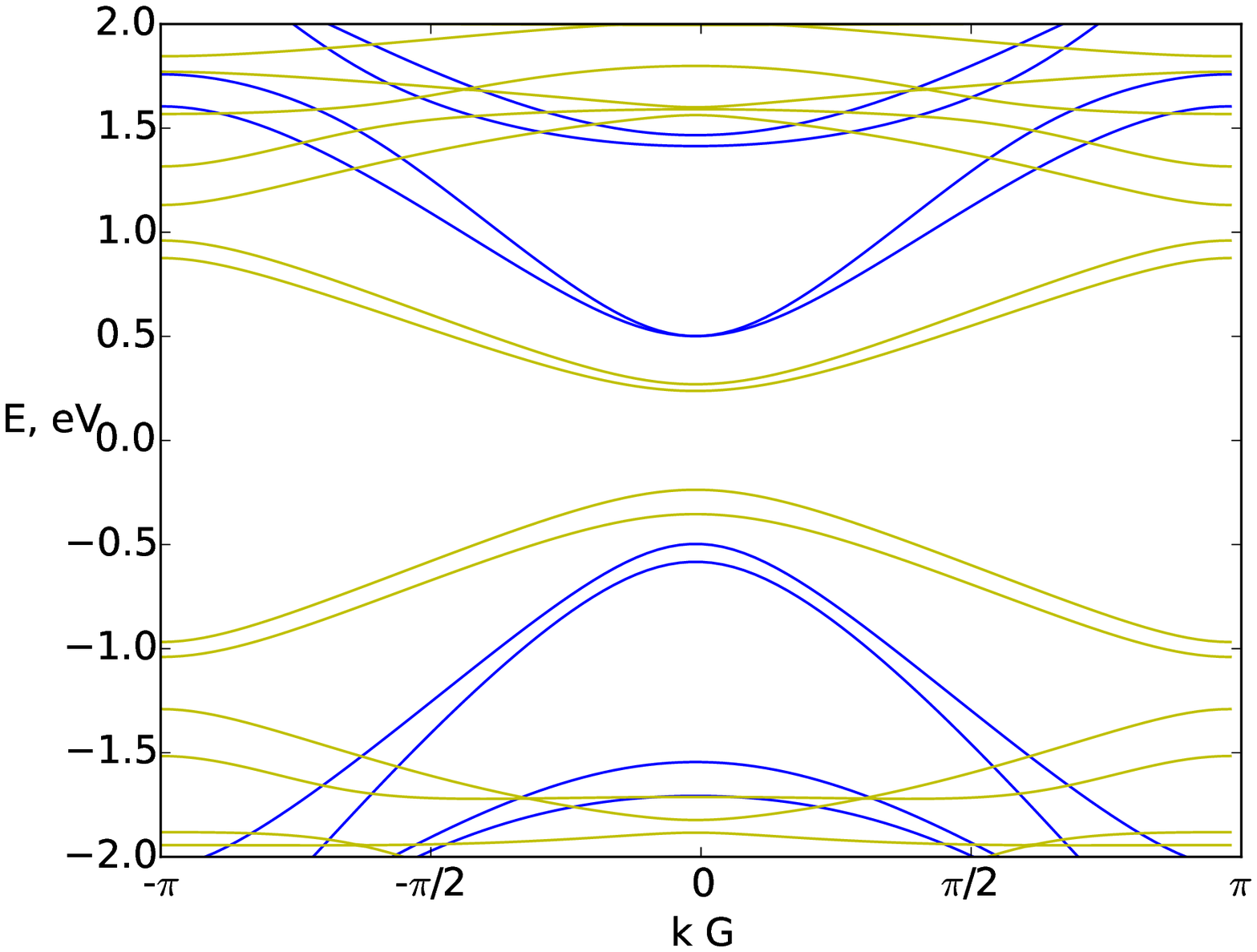} \label{fig1plus}
\caption{The electronic bandstructure for 6 atoms wide (left) and 8 atoms wide (right) structures, with antidot atoms left intact (blue curves) and removed (yellow curves):  energy (in eV) vs the dimensionless wavevector kG, where G is the spatial period of the structure. }
\end{figure*}
\begin{figure*}[htb]
\includegraphics*[width=.48\textwidth]{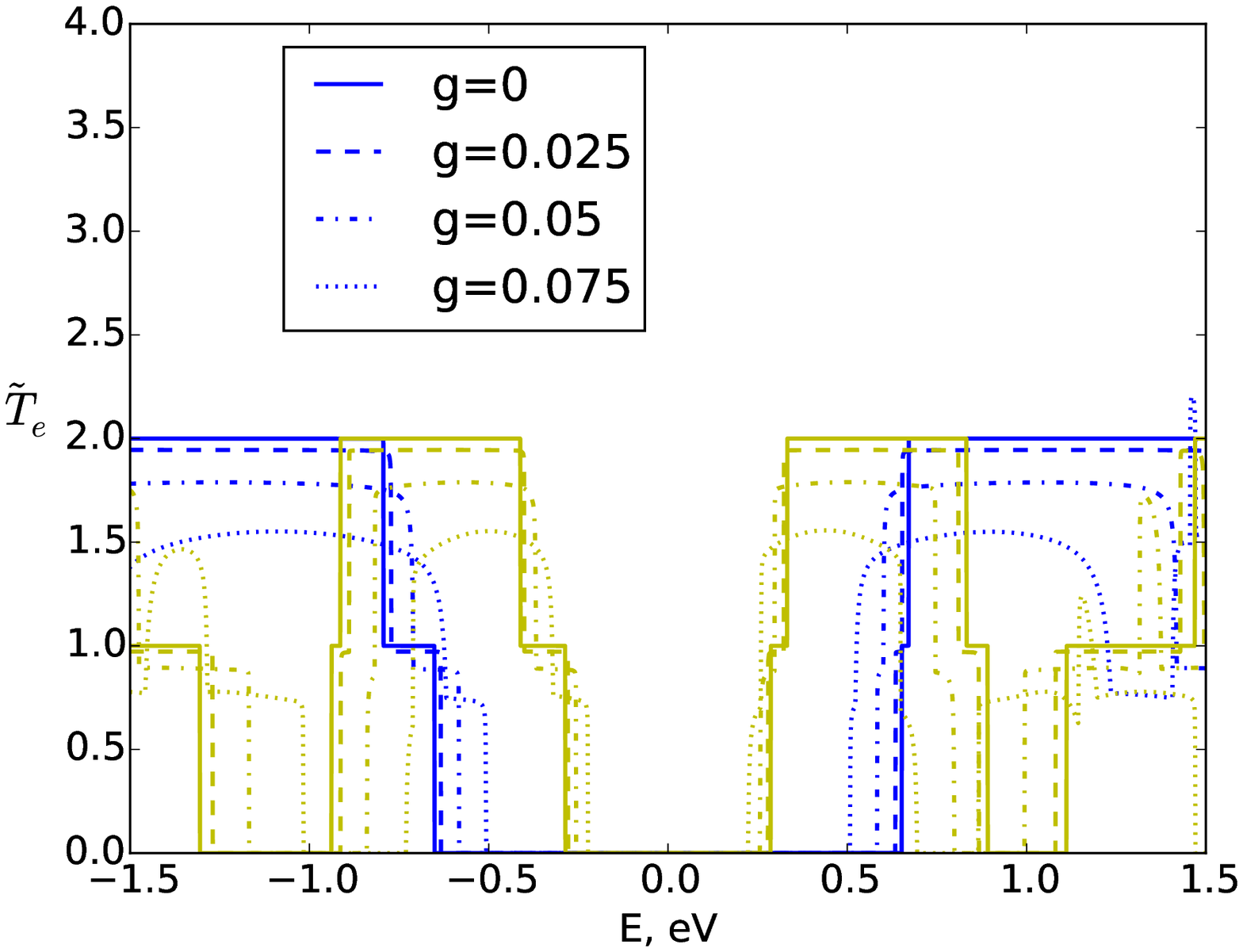} 
\includegraphics*[width=.48\textwidth]{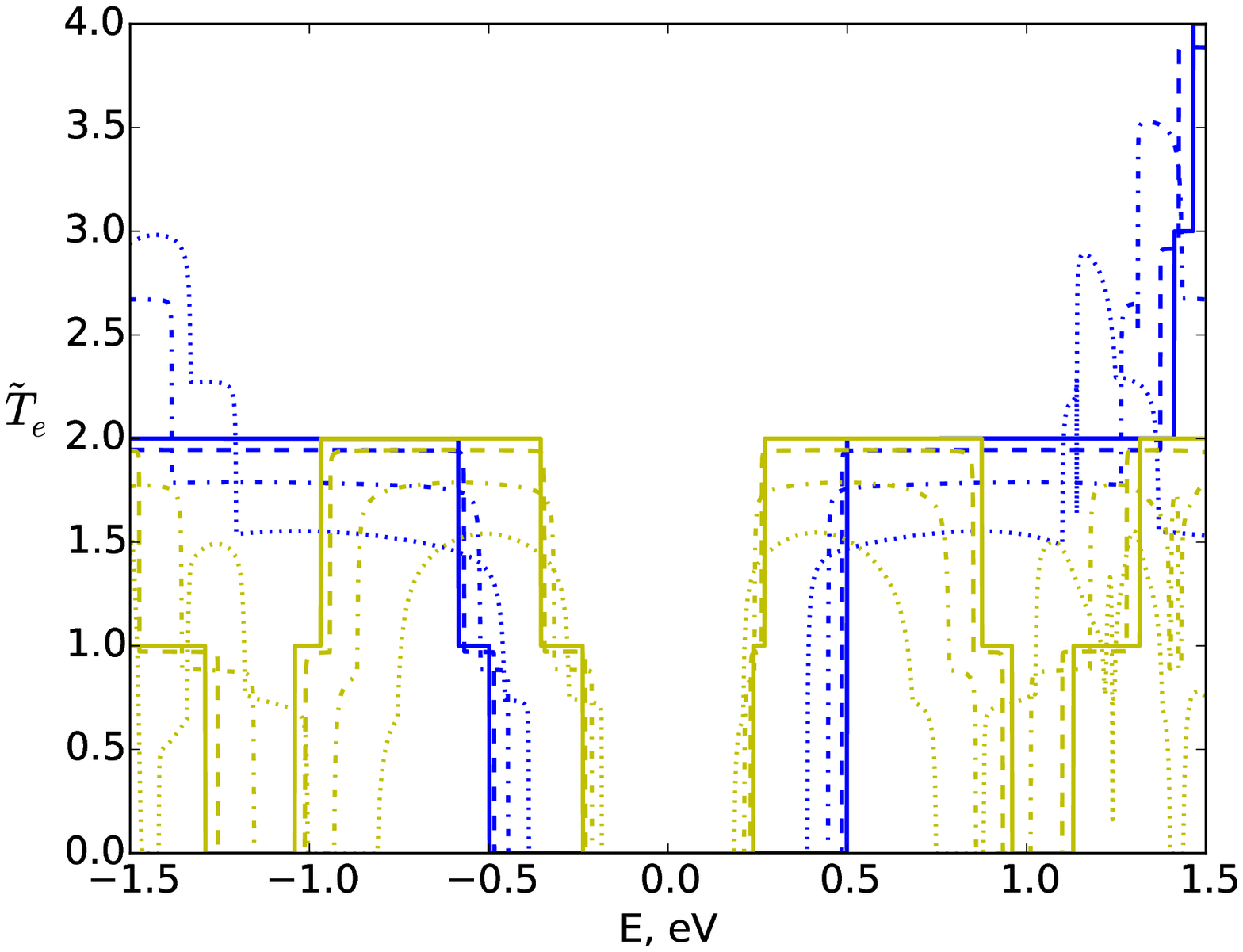} \label{fig2}
\caption{The effective electron transmission for 6 atoms wide (left) and 8 atoms wide (right) structures with antidot atoms left intact (blue curves) and removed (yellow curves).}
\end{figure*}
Note that the energy is measured from the center of the bandgap.  One should also note that one unit in Fig.3 corresponds to two spin channels for $\tilde{T}_e$  (cf. the difference between (\ref{Iesigma}) and (\ref{Ie})). For all the structures considered, the transmission function is found to be equal for both spin directions ($T_{e\sigma}=T_{e,-\sigma}$). The nearly degenerate energy levels at the edge of conduction band in Fig. 2 correspond to the sharp step of effective transmission function at the band edge energy in Fig. 3 for all the cases considered. The transmission function decreases with the increase of electron-phonon coupling. One can also see the smoothing of the transmission function at the band edges due to the electron-phonon interaction. 

The phonon transmission is shown in Fig.4.
\begin{figure*}[htb]
\includegraphics*[width=.48\textwidth]{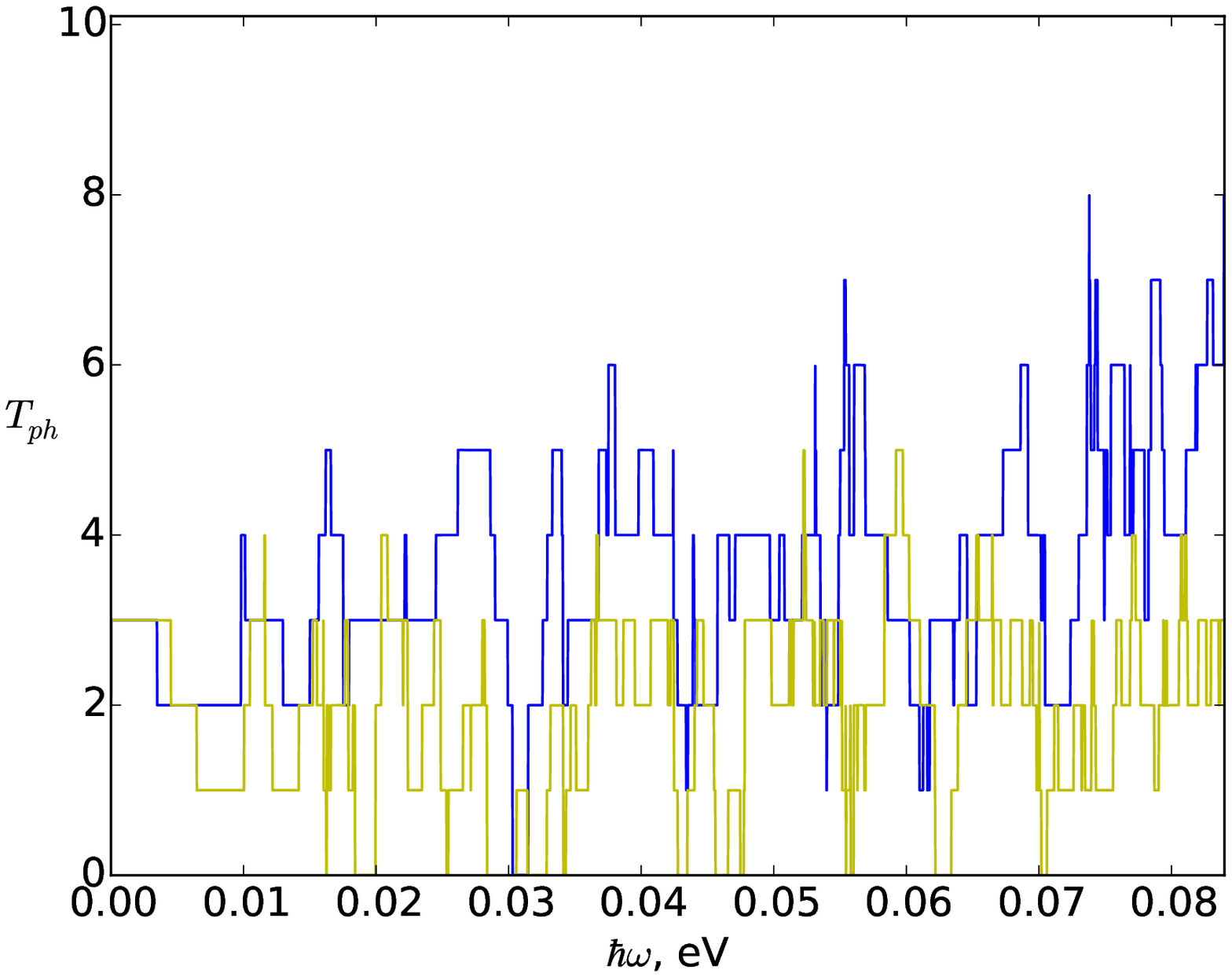}
\includegraphics*[width=.48\textwidth]{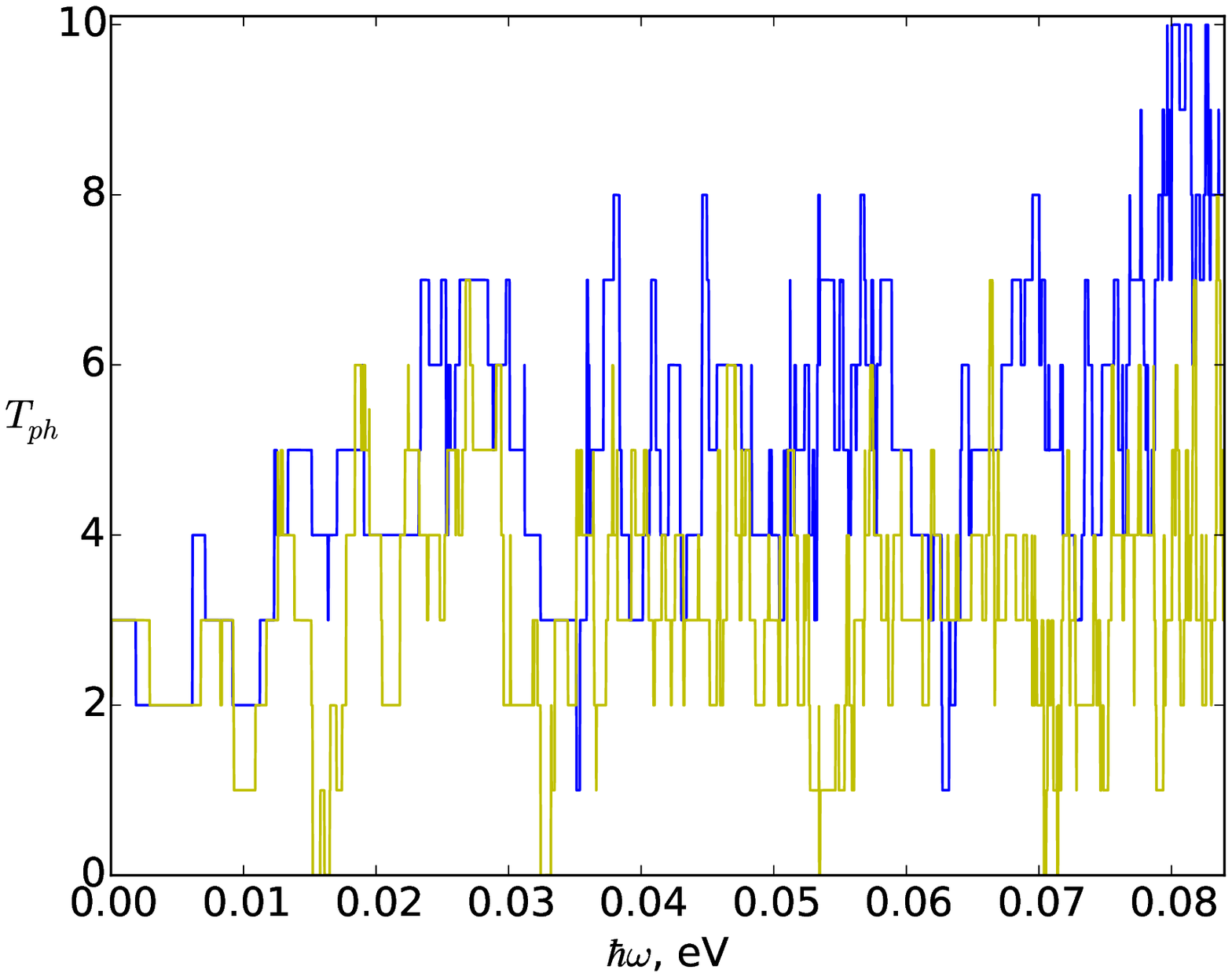} \label{fig3}
\caption{The phonon transmission function for 6 atoms wide (left) and 8 atoms wide (right) structures with antidot atoms left intact (blue curves) and removed (yellow curves).}
\end{figure*}
One can see the increase of number of phonon channels for the wider structure and the decrease of the number of channels for the antidot case. 

Fig.5 shows the thermopower.
\begin{figure*}[htb]
\includegraphics*[width=.48\textwidth]{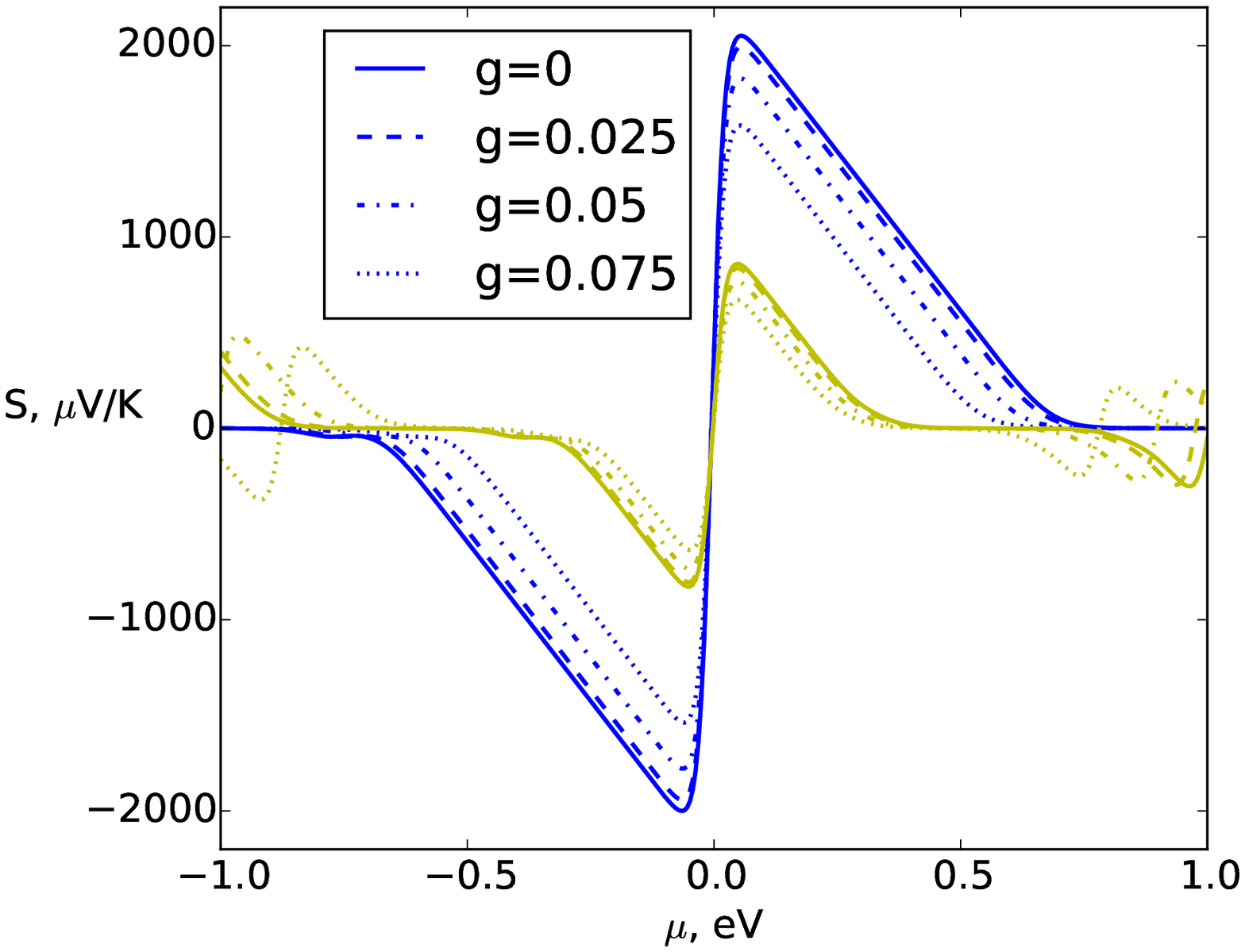}
\includegraphics*[width=.48\textwidth]{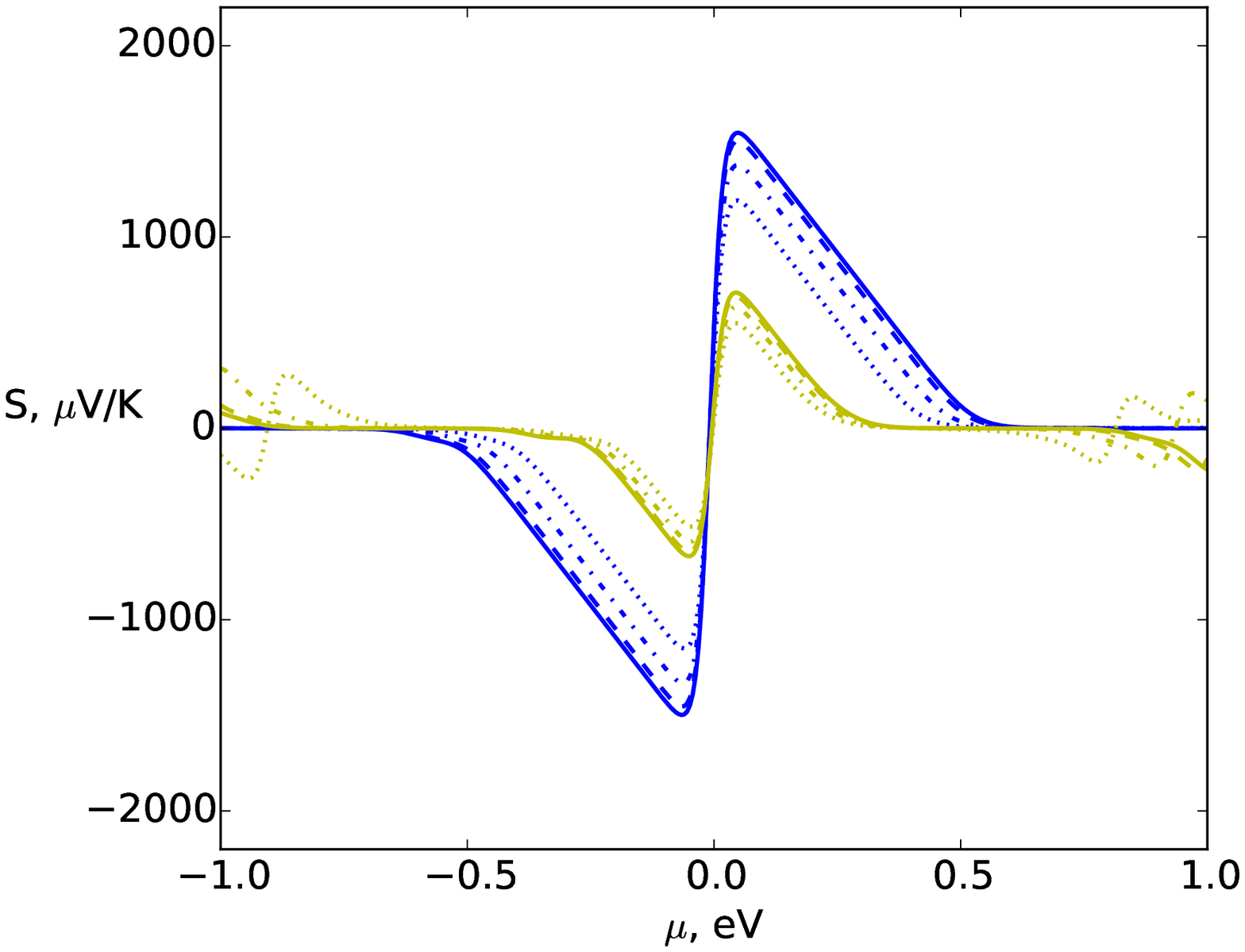} \label{fig3}
\caption{The thermopower for 6 atoms wide (left) and 8 atoms wide (right) structures with antidot atoms left intact (blue curves) and removed (yellow curves).}
\end{figure*}
The thermopower decreases with the increase of phonon coupling, and it's peaks shifts closer to the point $\mu=0$. This shift is more pronounced for the smaller secondary peaks at $\mu=\pm 1$ eV.

\begin{figure*}[htb]
\includegraphics*[width=.48\textwidth]{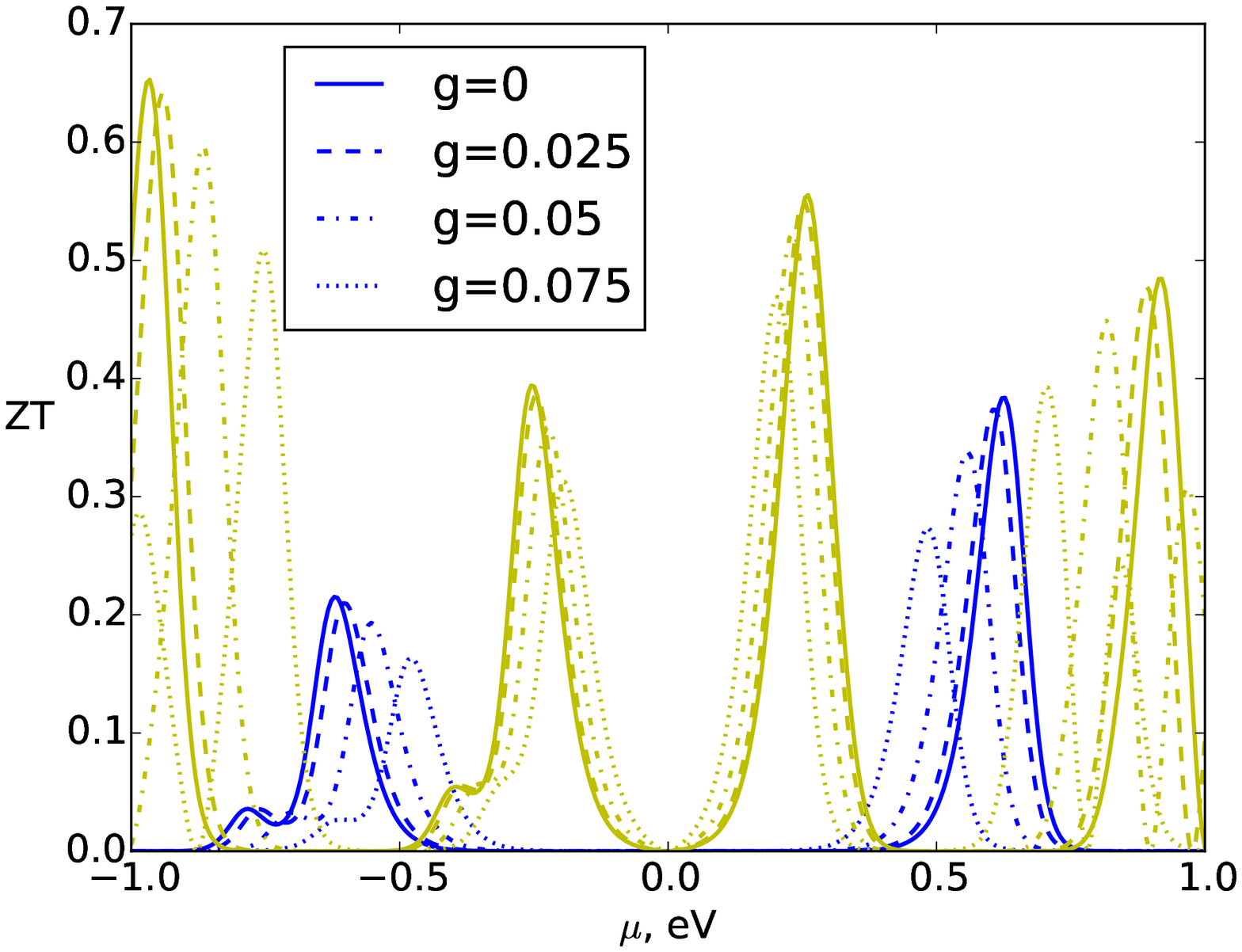}
\includegraphics*[width=.48\textwidth]{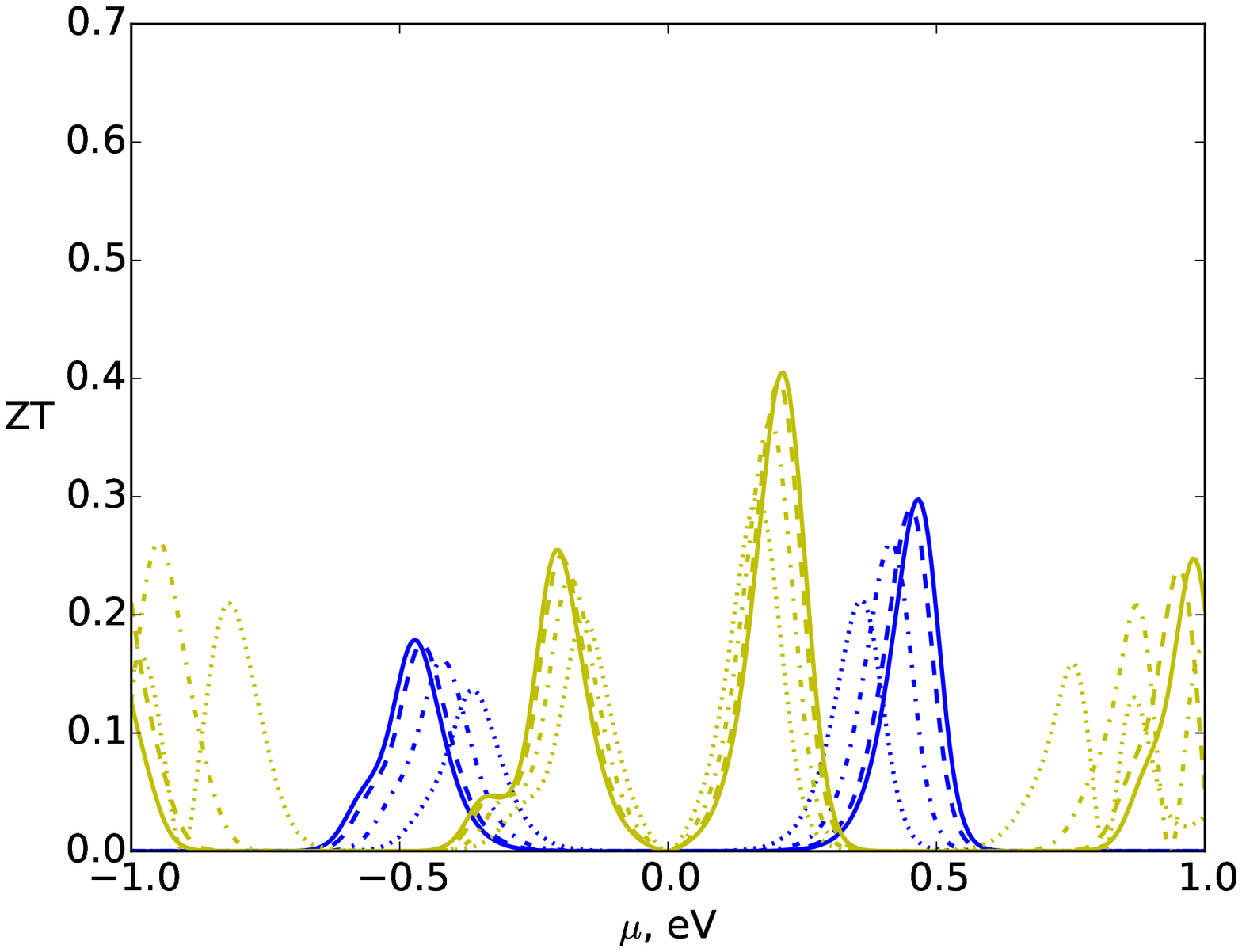} \label{fig3}
\caption{The figure of merit for 6 atoms wide (left) and 8 atoms wide (right) structures with antidot atoms left intact (blue curves) and removed (yellow curves).}
\end{figure*}
The thermoelectric efficiency is shown in Fig.6. As is seen, the peaks of ZT are reduced and shifted closer to the point $\mu=0$ with increasing $g$, similar to the case of the thermopower. For the antidot case the secondary peaks, which appear at the energies $\pm 1$ eV for the 6 atoms wide structure and $\pm$ 1.2 eV for the 8 atoms wide structure, have similar behavior. Both the decrease and the shift are more pronounced for the secondary peaks.  

\section{Conclusion}

We have investigated the influence of electron-phonon interaction on the thermoelectric properties of zigzag graphene nanoribbons, 6 and 8 atoms wide, with periodic edge vacancies. The structures with the period equal to three graphene lattice translations were investigated, and the influence of periodic antidots was studied.  The electron-phonon interaction was taken into account within the Hubbard-Holstein model, describing the density-density type interaction of the ZO-type optical phonon mode with conducting $\pi$-electrons, with the first and second nearest tight-binding terms included in the electronic Hamiltonian. We have used the Lang-Firsov transformation along with the Green's function method for the electronic Hamiltonian and 4th nearest neighbor approximation for the phonon dynamical matrix. The electron transmission functions and the thermoelectric figures of merit were calculated. 

We have found the appearance of peaks of figure of merit near the band edges of conduction and valence bands, as well as the secondary peaks in the case of periodic antidots. For all the structures and regimes considered, we have found the transmission function to be spin-independent. With the increase of electron-phonon coupling the modified electron transmission function decreases, its features shift toward the center of the band gap, and its step-like behavior smoothes out. This leads to the decrease of thermopower and thermoelectric efficiency, which is more pronounced for stronger electron-phonon coupling. Nevertheless, the peaks of thermoelectric efficiency that appear at the lower absolute values of chemical potentials remain dominant even at the highest coupling energies considered.

\end{document}